# ¿Tiene sentido limitar la coautoría científica? No existe inflación de autores en Ciencias Sociales y Educación en España

**Nicolas Robinson-Garcia[1] y Carlos B. Amat[2]**
[1] elrobinster@gmail.com ORCID: 0000-0002-0585-7359
[2] carbeam1@ingenio.upv.es ORCID: 0000-0002-6702-3317

INGENIO (CSIC-UPV), Universitat Politècnica de València

**Resumen**

Este trabajo analiza la evolución en coautoría en España en ciencias sociales para el periodo 2000-2013. El objetivo es explorar hasta qué punto están justificadas las limitaciones en número de autores que establecen las distintas agencias de evaluación españolas. El análisis de 11681 trabajos españoles de investigación en 20 categorías temáticas de las ciencias sociales revela que no hay inflación en el número de autores, que el tamaño de los equipos es homologable al de los trabajos internacionales en las mismas áreas y que el número de firmantes depende de la colaboración institucional y del nivel de internacionalización de los equipos. A excepción de los trabajos en Antropología y en Educación especial, el número de autores no es superior a cuatro. Sin embargo, son los trabajos con mayor número de autores los que obtienen un impacto mayor. El estudio evolutivo muestra una tendencia muy importante al alza en el tamaño de los equipos. En conjunto, estos resultados sugieren que carece de utilidad la limitación administrativa del número de firmantes que, con independencia de su eficacia en combatir la autoría honoraria, puede ir en perjuicio de la colaboración, la internacionalización y el posterior impacto de los trabajos.

**Palabras clave:** Bibliometría; evaluación científica; colaboración científica; Ciencias Sociales; autoría científica; España; coautoría

**Título en inglés:** Is it reasonable to limit scientific coauthorship? There is no inflation of co-authors in Social Sciences and Education in Spain

**Abstract**

This paper analyzes the evolution of coauthorship in Spain in the social sciences between 2000 and 2013. The goal is to explore to which extent limitations on the number of coauthors established by Spanish national evaluation agencies are justified. The analysis of 11681 papers authored by researchers affiliated to Spanish institutions in 20 subject categories of the social sciences reveals that there is no inflation in the number of authors, team size is similar to that found in foreign papers and the number of authors is dependent on international and institutional collaboration. With the exception of Anthropology and Special Education, in no area the average of authors by paper is higher than four. However, papers with a higher number of authors receive more citations. Overall, our results suggest that there is no justification on limiting the number of coauthors in publications, acknowledging that their inclusion in the criteria employed by Spanish evaluation agencies is to prevent honorary authors. Such limitation endangers institutional and international collaboration, and consequently, high impact research.

## INTRODUCCIÓN

La evaluación de la actividad investigadora es una constante en la vida del investigador español. Acreditaciones, sexenios y demás procesos evaluativos, hacen que el científico centre gran parte de sus esfuerzos investigadores en producir resultados que se ajusten a estrictos criterios cuantitativos establecidos por las agencias nacionales de evaluación (Rafols y otros, 2016). En los últimos años se observa una creciente sistematización de criterios mediante la inclusión de indicadores cuantitativos en los procesos evaluativos de promoción y acceso a la carrera investigadora y universitaria (Ruiz-Pérez y otros, 2010). En los criterios que establecen las dos agencias estatales encargadas de dichos procesos evaluativos (la Comisión Nacional para la Evaluación de la Actividad Investigadora conocidas por sus siglas CNEAI, y la Agencia Nacional de Evaluación de la Calidad y la Acreditación, ANECA), se muestra un genuino interés por el número de autores que los trabajos evaluados tienen, incentivando al evaluado a ocupar siempre posiciones de liderazgo en el orden de firma. Ejemplo de ello son los criterios para los tramos de investigación que establece la CNEAI (Boletín Oficial del Estado, 2016, sec. III) donde en cada uno de los 11 campos disciplinares en los que se dividen los comités evaluadores se hace mención expresa a la autoría, señalando la necesidad de justificar la coautoría (Cabezas-Clavijo y Torres-Salinas, 2015).

## ANTECEDENTES

El estudio de la colaboración, un elemento esencial del proceso investigador, representa uno de los principales frentes de investigación dentro del ámbito de la bibliometría. La cooperación entre científicos permite plantear objetivos de investigación ambiciosos e inabordables si se acometieran en solitario (Katz y Martin, 1997). Por otra parte, posibilita la diseminación de los resultados, y juega un papel relevante en la formación de jóvenes investigadores y en la transmisión rápida del conocimiento generado (Wray, 2002). La colaboración científica es reflejo y consecuencia de los cambios producidos en las dinámicas de funcionamiento del sistema científico de los últimos 30 años, donde la colaboración ha ganado cada vez más peso, siendo actualmente minoritarios los trabajos escritos por un solo autor (Wuchty y otros, 2007). Esta transformación hacia el trabajo colaborativo también se ha visto reflejada en aquellos trabajos considerados de mayor calidad o impacto científico. Así pues, encontramos que también son mayoría los trabajos colaborativos dentro de los altamente citados (Wuchty y otros, 2007). Del mismo modo, aunque se asocia un mayor impacto a trabajos en colaboración internacional, también se observa cierta ventaja en la citación para trabajos en colaboración entre autores de una misma institución o de distintas instituciones asociadas al mismo país de procedencia (Katz y Hicks, 1997).

En el caso específico de las disciplinas de las Ciencias Sociales, la tendencia en el número de autores por trabajo, también es creciente, a pesar de presentar un ritmo más lento que otras áreas (Henriksen, 2016). Un crecimiento también reflejado en el caso de España (De Filippo y otros, 2014). La explicación en dicho incremento no sólo se encuentra en el aumento de la colaboración internacional en estas disciplinas, sino también en algunas de las áreas asociadas a las Ciencias Sociales en el análisis de grandes sets de datos y la aplicación de metodologías de campo y de análisis de datos (Henriksen, 2016) que requieren de equipos con miembros especializados y en los que se requiere una distribución de tareas (Larivière y otros, 2016).

La coautoría científica es sólo un reflejo parcial de la colaboración entre investigadores. Por tanto, no todos aquellos que participan en un estudio de investigación acaban apareciendo en el listado de autores. Sin embargo, los criterios que siguen los investigadores para determinar quién aparece en cada trabajo pueden variar desde la inclusión de cualquier individuo que haya participado en el estudio independientemente de su grado de implicación (Stokes y Hartley, 1989), hasta la exclusión de cualquier colaborador que se encuentre jerárquicamente en una posición inferior, a pesar de haber sido claves para el desarrollo del estudio (Shapin, 1989). Quién aparece o no al final en el listado de autores, suele ser el resultado de negociaciones en ocasiones complejas y controvertidas. Por tanto, la elección de autores es una actividad social, y como tal, no está exenta de malas prácticas (Bosch, 2011). Todo ello, dificulta discernir mediante métodos bibliométricos hasta qué punto queda justificado el número de autores que presenta un trabajo y la contribución al mismo de cada uno de ellos.

Es más, cabe cuestionarse, en qué medida tiene sentido analizar y evaluar el trabajo de unidades de análisis tan específicas (como es el investigador) en lugar de revisar los métodos de monitorización y evaluación, cuando resulta cada vez más difícil determinar 'de quién' son los trabajos que producen estos investigadores de manera colaborativa (Hicks y Katz, 1996).

La co-autoría de trabajos científicos ha suscitado el interés de investigadores españoles en repetidas ocasiones. Quizá la más destacada haya sido el congreso "La colaboración científica: una aproximación multidisciplinar", que ofreció 47 comunicaciones sobre el tema (Aguiló Calatayud y otros, 2013) algunas publicadas luego en un número monográfico de esta misma revista (vol. 37 num. 4 de 2014). Por otra parte, además de algunos trabajos episódicos y meramente cuantitativos, como el de Yegros-Yegros y otros (2012), destacan las iniciativas del Observatorio IUNE (http://www.iune.es/es_ES) y del Grupo EC3 de la Universidad de Granada. El primero, que depende del instituto interuniversitario "Investigación Avanzada sobre Evaluación de la Ciencia y la Universidad" mantiene un sistema interactivo de análisis de la actividad investigadora de las universidades españolas que, entre otros indicadores, ofrece un "índice de coautoría", el promedio de autores por documento producido por cada universidad. De forma paralela, el grupo EC3 desarrolló el portal Co-author Index (http://www.coauthorindex.info) que ofrece los principales estadísticos descriptivos de las distribuciones del número de autores de los artículos españoles de investigación.

El presente trabajo analiza las posibles consecuencias que puede tener establecer limitaciones en la coautoría de cara a la producción y al impacto de trabajos de investigación. Para ello, analizamos la evolución de la coautoría de trabajos españoles para el periodo 2000-2013 en Ciencias Sociales y Educación con el objetivo de analizar dos variables que consideramos pueden verse afectadas por este tipo de políticas: 1) la colaboración institucional e internacional y 2) la llamada investigación de excelencia (definida a través de la citación). Utilizando como referente los criterios (provisionales) para la acreditación a los cuerpos de titular y catedrático de universidad regulados por el programa ACADEMIA de la ANECA [1] publicados el pasado mes de noviembre, centramos nuestro estudio en las áreas de Ciencias Sociales y Educación, siguiendo la estructuración de disciplinas de los 21 paneles establecidos por la ANECA. En estos dos paneles se fija un umbral de un máximo de cuatro firmantes por trabajo y se indica que se penalizarán trabajos con un mayor número de autores.

---

[1] Los nuevos criterios de la ANECA fueron publicados el pasado 14 de noviembre, 2016 y están disponibles en la web: http://www.mecd.gob.es/servicios-al-ciudadano-mecd/catalogo/general/educacion/academia/ficha/academia.html

Para ello, el trabajo se estructura del siguiente modo. En primer lugar, hacemos una breve revisión bibliográfica sobre la importancia de la colaboración científica y el uso de la autoría científica como herramienta de reconocimiento científico en los procesos evaluativos. En segundo lugar, describimos el proceso de colección y procesamiento de datos, así como la metodología presentada. Tras presentar los resultados del estudio, discutimos en función de los objetivos planteados, la pertinencia o no de establecer política de evaluación investigadora que establezcan el número de autores óptimo que deben tener los trabajos científicos.

## DATOS Y MÉTODOS

### Recolección y procesamiento de datos

Este trabajo analiza la producción de autores afiliados a instituciones españolas en Ciencias Sociales y Educación durante el periodo 2000-2013. Para ello, analizamos un total de 11681 publicaciones extraídas de la colección principal del Web of Science. Todas proceden del Social Sciences Citation Index y corresponden a documentos citables (artículos, revisiones y cartas). Se han recuperado empleando un requisito inicial: la presencia de alguna institución española (CU=Spain) en el campo de direcciones. La descarga de trabajos se ha hecho anualmente entre los meses de abril y mayo desde el año 2000.

Hemos procesado las 56 categorías temáticas (WC) asignadas a los trabajos. La ANECA establece 21 paneles para los cuáles enuncia unos criterios específicos. Dentro del ámbito de las Ciencias Sociales y Jurídicas incluye cinco paneles (Derecho, Ciencias de la Educación, Ciencias del Comportamiento y Ciencias Sociales). Los paneles de Ciencias Sociales y de Educación indican de manera explícita una penalización para aquellos trabajos que se presenten con más de cuatro autores. A fin de analizar específicamente lo que ocurre en las disciplinas evaluadas por estos dos paneles, se revisaron las categorías temáticas desechando aquellas que estuvieran cubiertas por otro panel.

Se han individualizado y procesado los autores individuales (AU) y se han descartado los autores colectivos (CA) de cada trabajo; asimismo se han procesado las instituciones que han contribuido a cada trabajo. Los registros posteriores a 2007 relacionan a cada autor con su institución y país de procedencia (C1). Esto nos ha permitido diferenciar los autores nacionales de los extranjeros. La distinción entre nacional y extranjero se emplea únicamente para identificar trabajos de colaboración nacional e internacional, pero no para el cálculo de los indicadores de promedio de autores por trabajo. A continuación se describen los indicadores empleados:

**Promedio de autores por trabajo**. Este indicador se ofrece por categoría temática y para cada trabajo en el que al menos una de las instituciones firmantes sea española.

**Tasa de variación.** Se emplea para analizar la tendencia creciente o decreciente en el promedio de autores por categoría temática. En este caso, se ha calculado en función del promedio de autores en el quinquenio 2000-2004 y el quinquenio 2009-2013.

**Promedio de citas (PCIT).** Hemos establecido una ventana de citación de tres años para cada trabajo a fin de que no haya un sesgo determinado por los trabajos más antiguos

**Impacto normalizado (FNCS).** Para su cálculo se ha empleado el cálculo descrito por Lundberg (2007), tomando como valor de referencia de cada campo temático la media de citas recibidas por los trabajos españoles, y no la correspondiente a los trabajos a nivel mundial.

## Indicadores y análisis estadístico

Tras hacer un análisis descriptivo de la evolución temporal en el número de autores medio por categoría temática y diferencias de citación según el número de autores y tipo de colaboración (interinstitucional tanto a nivel nacional como internacional), incluimos una serie de análisis estadísticos para analizar la relación entre el número de autores y la colaboración internacional, así como su posible influencia en el impacto de los trabajos. Para estudiar la relación entre colaboración internacional y número de autores realizamos una regresión múltiple para cada categoría temática. Los resultados se ofrecen en la tabla II. Para analizar la relación entre el impacto y el número de autores, empleamos dos indicadores: número de citas e impacto normalizado. Dividimos los trabajos en dos grupos: aquellos con más de cuatro autores y aquellos otros con cuatro o menos. El primer grupo comprende un total de 1787 trabajos y el resto son 9894. Aplicamos un ANOVA empleando la prueba de Kruskal-Wallis, un método no paramétrico que asume la no normalidad de las distribuciones. Compara medianas entre dos grupos de datos, asumiendo que ambas distribuciones son similares. El objetivo de este análisis es de identificar diferencias en la citación en función del número de autores. En este caso, el análisis se hizo para el set de datos completo.

Hemos incluido material complementario disponible en Amat y Robinson-Garcia (2017). Para consultar las tablas S1-S4 se ruega al lector consulte dicho material.

## RESULTADOS

España publicó un total de 11681 trabajos en las 20 categorías temáticas incluidas en los campos de Ciencias Sociales y Educación. El número de autores promedio para esta población es de 2,9. La tabla I muestra la evolución temporal en el número de autores promedio por categoría temática. En dos áreas de investigación, Antropología y Educación Especial, parece ser norma un equipo mayor de cuatro autores. Pero, más allá de algunos artefactos, como el promedio superior a 19 autores en Comunicación de 2003 (debido a un único trabajo con más 100 autores) o de los repuntes episódicos en el tamaño de los equipos en otras disciplinas, no observamos inflación en el número de autores. Ahora bien, el tamaño es un concepto estático y, para poder comparar adecuadamente las cifras observadas, conviene atender al crecimiento, un concepto longitudinal, dinámico. La variación porcentual entre la media de autores por trabajo en cada disciplina en el quinquenio 2000-2004 y el periodo 2009-2013 muestra una tendencia positiva para todas las disciplinas. Es decir, se observa un crecimiento paulatino en la coautoría científica. La única excepción es Comunicación, debido a la distorsión que produce el trabajo anteriormente mencionado. Se observa un incremento medio del 27% (30% si no se considera Comunicación). En 7 de las 20 categorías observadas, el incremento supera el 30% y es superior al 70% en Relaciones Internacionales (73).

La variación en el número de colaboradores es paralela a número de instituciones (tabla S1) donde la tasa de variación es positiva para 18 categorías temáticas con la excepción de Estudios Étnicos (-11,2) y Turismo, Deporte y Ocio (-6,4). Este crecimiento se debe en parte a la mayor colaboración entre distintas instituciones españolas (tabla S2). De hecho, el porcentaje de variación entre los quinquenios inicial y final supera el 30% en 16 de las 20 categorías disciplinarias y en Relaciones laborales se duplica, pasando de 0,77 a 1,57 instituciones españolas por trabajo. La proporción de trabajos con participación de instituciones extranjeras (tabla S3) ha pasado de un 25% a un 35%. Destacando, Relaciones Internacionales que pasa de un promedio de 0,08 a 1,13 y Educación especial (de 0,34 a 1,6).

**Tabla I. Producción española en Ciencias Sociales y Educación y estadísticos descriptivos de autores e instituciones por año según el SSCI. Periodo 2000-2013. En rojo años en los que la media de autores es mayor a 4 autores por trabajo.**

| Categoría WoS | 2000-2004 | 2009-2013 | var % | # pubs. | 2000-2013 |
|---|---|---|---|---|---|
| Anthropology | 3,9 | 5,0 | 28,3 | 1012 | |
| Area Studies | 1,8 | 2,2 | 18,1 | 52 | |
| Communication | 6,7 | 2,9 | -56,2 | 562 | |
| Education & Educational Research | 2,8 | 3,0 | 8,3 | 2484 | |
| Education, Scientific Disciplines | 3,2 | 4,0 | 24,2 | 1289 | |
| Education, Special | 3,7 | 5,2 | 39,7 | 157 | |
| Ergonomics | 3,4 | 3,8 | 13,2 | 320 | |
| Ethnic Studies | 1,9 | 2,3 | 20,0 | 34 | |
| Family Studies | 2,9 | 3,9 | 32,6 | 160 | |
| Hospitality, Leisure, Sport & Tourism | 2,4 | 2,9 | 22,9 | 457 | |
| Industrial Relations & Labor | 2,3 | 3,0 | 30,3 | 185 | |
| Information Science & Library Science | 2,6 | 3,1 | 18,3 | 1696 | |
| International Relations | 1,7 | 2,9 | 73,0 | 294 | |
| Linguistics | 2,1 | 2,6 | 23,8 | 998 | |
| Political Science | 1,7 | 2,2 | 24,7 | 528 | |
| Social Issues | 2,5 | 3,2 | 26,8 | 174 | |
| Social Sciences, Interdisciplinary | 2,7 | 3,5 | 30,9 | 736 | |
| Social Sciences, Mathematical Methods | 2,1 | 2,7 | 25,8 | 959 | |
| Social Work | 2,5 | 3,6 | 44,1 | 130 | |
| Sociology | 2,1 | 2,8 | 37,3 | 738 | |

Estos incrementos en la autoría y el tipo de colaboración institucional no se observan al agregar los trabajos de todas las categorías, debido a las grandes diferencias existentes entre ellas. Esto se refleja en la figura 1, donde el patrón de promedio de autores y año no es tan evidente como sugiere la tasa de variación. Así, se observa que en sólo dos años el promedio de trabajos en colaboración internacional sobrepasó los cuatro autores fijados por la ANECA. En el caso de 2003, esta subida en el promedio que los cinco autores por trabajo, se debe a un caso aislado. El trabajo titulado 'Are men universally more dismissing than women? Gender differences in romantic attachment across 62 cultural regions' cuenta con más 100 autores. La colaboración entre autores de distintas instituciones así como la colaboración internacional muestran un patrón estable con un leve incremento en el promedio de autores para trabajos firmados por una sola institución.

En cualquier caso, este promedio sigue siendo inferior, nunca alcanzando los tres autores por trabajo.

Con el fin de analizar la relación entre número de coautores y la colaboración institucional, realizamos una regresión múltiple (Tabla II) en la que se analiza la influencia del número de instituciones extranjeras y nacionales en el número de autores de los trabajos. En ella se observa cómo el número de autores se asocia más con el número de instituciones extranjeras que con el número de instituciones españolas que colaboran en cada trabajo. No obstante, esta relación permite predecir sólo el 47 por ciento de los casos. Por categoría temática, el modelo más ajustado se observa para Comunicación ($R^2$=0,89), Política (0,7), Educación especial (0,64), Relaciones internacionales (0,6), Antropología (0,59), Problemática social (0,52) y Estudios de familia (0,51).

**Figura 1. Distribución del número de instituciones (A) y autores por trabajo y año. Periodo 2000-2013. Línea roja discontínua marca el límite de cuatro autores a partir del cuál penaliza la ANECA.**

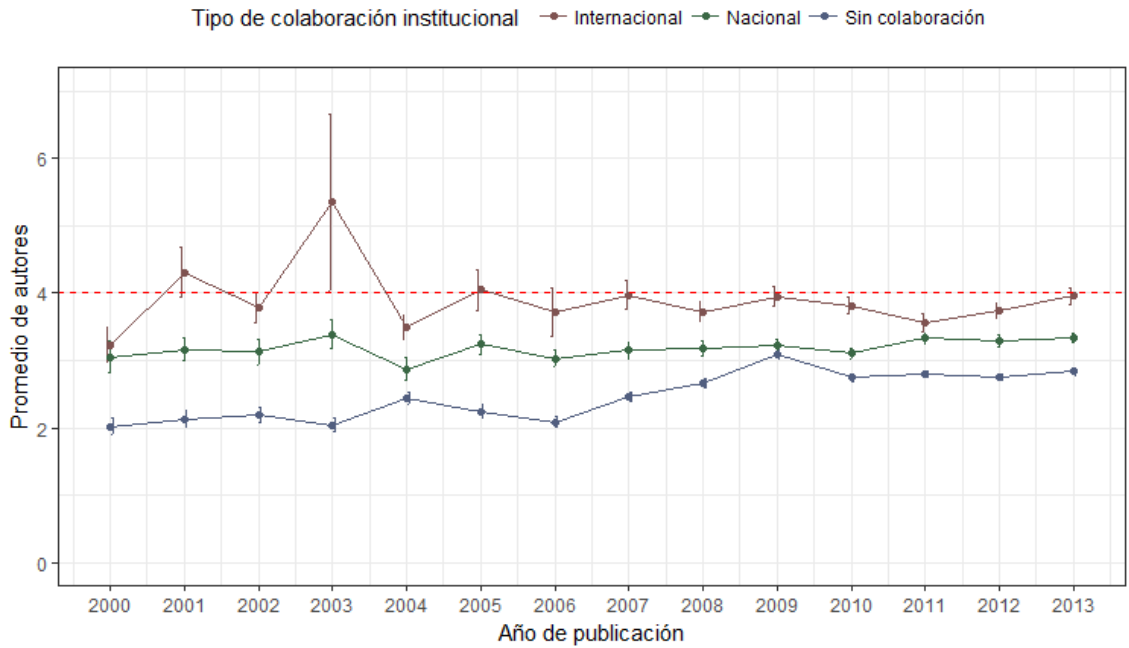

**Tabla II. Influencia del tipo de colaboración institucional (nacional vs. Internacional). Regresión múltiple por categoría temática. Intervalo de confianza en paréntesis. $*p < 0.5$, $**p < 0.05$, $***p < 0.01$.**

| Categorías temáticas SSCI | Colaboración internacional | Colaboración Nacional | $R^2$ justada | F |
|---|---|---|---|---|
| Anthropology | 1,02*** (0,96-1,08) | 0,99*** (0,89 -1,1) | 0,59 | 730,50 |
| Area Studies | 0,2* (-0,1-0,5) | 0,54** (-0,06- 1,14) | 0,05 | 2,33 |
| Communication | 1,67*** (1,63-1,72) | 0,83 | 0,89 | 2384,00 |
| Education & Educational Research | 0,89*** (0,8-0,94) | 0,51***(0,43-0,6) | 0,22 | 348,90 |
| Education, Scientific Disciplines | 0,66*** (0,55-0,76) | 0,52*** (0,40-0,65) | 0,14 | 104,60 |
| Education, Special | 1,03*** (0,9-1,15) | 0,95*** (0,66-1,24) | 0,64 | 141,60 |
| Ergonomics | 0,76*** (0,58-0,93) | 0,92***(0,79-1,05) | 0,43 | 121,30 |
| Family Studies | 0,88*** (0,68-1,07) | 1,15*** (0,94-1,36) | 0,51 | 84,21 |
| Hospitality, Leisure, Sport & Tourism | 0,51*** (0,37-0,65) | 0,72*** (0,58-0,86) | 0,23 | 69,16 |
| Industrial Relations & Labor | 0,56*** (0,43-0,7) | 0,42*** (0,18-0,65) | 0,27 | 35,05 |
| Information Science & Library Science | 0,44*** (0,37-0,51) | 0,52*** (0,45-0,6) | 0,15 | 154,50 |
| International Relations | 1,07*** (0,96-1,17) | 0,88*** (0,56-1,21 | 0,60 | 222,90 |
| Linguistics | 0,93*** (0,85-1) | 0,73*** (0,62-0,85) | 0,40 | 340,20 |
| Political Science | 1*** (0,94-1,05) | 0,4*** (0,26-0,55) | 0,70 | 626,30 |
| Social Issues | 0,84*** (0,71-0,98) | 0,73*** (0,48-1) | 0,52 | 94,10 |
| Social Sciences, Interdisciplinary | 0,78*** (0,69-0,86) | 0,77*** (0,63-0,88) | 0,38 | 228,20 |
| Social Sciences, Mathematical Methods | 0,43*** (0,37-0,48) | 0,45*** (0,38-0,53) | 0,24 | 149,60 |
| Social Work | 0,61*** (0,37-0,85) | 0,7*** (0,42-1) | 0,24 | 21,79 |
| Sociology | 0,88*** (0,81-0,96) | 0,78*** (0,65-0,9) | 0,46 | 312,30 |

Por último, analizamos la relación entre colaboración e impacto. La tabla III muestra los indicadores de impacto por categoría temática. La tabla S4 muestra el número de trabajos sobre los que se calculan dichos indicadores. En líneas generales, observamos como el promedio de citas para trabajos con más de cuatro autores duplica el promedio de citas del resto de trabajos. De hecho, el impacto normalizado de estos trabajos es un 50% mayor para los trabajos con más de cuatro autores. La categoría temática con mayor promedio de citas por trabajo e impacto normalizado es Antropología, seguido de Educación Especial. Los trabajos con más de cuatro autores tienen mayor impacto científico en todas las categorías temáticas salvo en el caso de Turismo, Deporte y Ocio (0,8 de diferencia en el promedio de citas y 0,4 de diferencia para el impacto normalizado) y moderadamente inferior para Ergonomía (0,3 y 0,2 puntos de diferencia para el promedio de citas e impacto normalizado respectivamente) Biblioteconomía y Documentación (0,1 puntos menos para ambos indicadores de impacto) y Estudios de la Familia (una diferencia a favor de los trabajos con más de cuatro autores de 0,1 para el promedio de citas y 0,1 puntos negativa según el impacto normalizado).

**Tabla III. Promedio de citas (PCIT) e impacto normalizado (FNCS) por categoría para todos los trabajos, trabajos con más de cuatros autores y con 4 o menos autores. Se excluye el cálculo para categorías con menos de 18 trabajos con más de cuatro autores.**

| Categorías temáticas SSCI | Total | | Pubs con > 4 autores | | Pubs con ≤ 4 autores | |
|---|---|---|---|---|---|---|
| | PCIT | FNCS | PCIT | FNCS | PCIT | FNCS |
| Anthropology | 3,6 | 1,0 | 4,7 | 1,3 | 2,8 | 0,8 |
| Area Studies | 1,1 | 1,3 | — | — | — | — |
| Communication | 1,7 | 1,3 | 3,7 | 3,0 | 1,4 | 1,1 |
| Education & Educational Research | 1,1 | 1,0 | 2,0 | 1,6 | 1,0 | 0,9 |
| Education, Scientific Disciplines | 1,2 | 1,1 | 1,7 | 1,7 | 1,1 | 0,9 |
| Education, Special | 2,8 | 1,0 | 3,7 | 1,2 | 2,2 | 0,8 |
| Ergonomics | 2,2 | 1,0 | 2,0 | 0,8 | 2,3 | 1,0 |
| Ethnic Studies | 1,6 | 1,1 | — | — | — | — |
| Family Studies | 2,5 | 1,0 | 2,6 | 1,0 | 2,5 | 1,1 |
| Hospitality, Leisure, Sport & Tourism | 2,6 | 1,1 | 1,8 | 0,7 | 2,6 | 1,1 |
| Industrial Relations & Labor | 1,4 | 1,0 | — | — | — | — |
| Information Science & Library Science | 2,1 | 1,0 | 2,0 | 0,9 | 2,1 | 1,0 |
| International Relations | 2,1 | 1,2 | 7,5 | 3,0 | 1,7 | 1,0 |
| Linguistics | 1,8 | 1,3 | 4,8 | 3,5 | 1,5 | 1,1 |
| Political Science | 1,7 | 1,7 | — | — | — | — |
| Social Issues | 2,0 | 1,1 | 2,8 | 2,3 | 1,9 | 0,9 |
| Social Sciences, Interdisciplinary | 1,7 | 1,7 | 2,3 | 1,3 | 1,6 | 0,9 |
| Social Sciences, Mathematical Methods | 1,9 | 1,1 | 3,1 | 1,5 | 1,9 | 1,1 |
| Social Work | 1,9 | 1,0 | 2,2 | 1,3 | 1,9 | 1,0 |
| Sociology | 1,6 | 1,2 | 3,5 | 2,2 | 1,5 | 1,1 |
| **Total** | **1,9** | **1,1** | **3,0** | **1,5** | **1,6** | **1,0** |

Las mayores diferencias a favor de los trabajos con más de cuatro autores se observan en Relaciones internacionales, con un promedio de citación más de cuatro veces superior y tres veces superior en el caso del impacto normalizado. En segundo lugar, destaca Lingüística, que triplica para ambos indicadores el impacto de los trabajos con más de cuatro autores a aquellos con cuatro o menos. Por último, destaca Comunicación con un promedio de citas 2,6 superior y un impacto normalizado 2,7 veces superior, otra vez a favor de los trabajos con más de cuatro autores.

Desde 2007, Web of Science asocia cada autor y su filiación institucional. Los datos, sin embargo, son completos y correctos sólo a partir de 2008. Hemos podido diferenciar los autores nacionales de los extranjeros en los 8168 trabajos publicados entre 2008 y 2013 en las áreas temáticas que analizamos. En el 70% de los trabajos no existen autores afiliados a instituciones extranjeras. Las proporciones de casos en que el número de autores nacionales iguala al de extranjeros ronda el 10%. También rondan esa cifra la proporción de trabajos en los que predominan los autores nacionales. En los trabajos con contribución internacional, la proporción media de autores extranjeros en todos los trabajos analizados se ha mantenido ligeramente por encima de la mitad; en 2008 era del 50,7 % y en 2013 del 52,4 %.

En relación al análisis ANOVA Kruskall-Wallis, que compara los indicadores de impacto para trabajos con más de 4 autores y con 4 o menos, observamos que los trabajos con más autores han recibido una media de casi 3 citas, la mitad ha sido citados 2 o más veces y una cuarta parte de los trabajos han recibido más de 4 citas. Su impacto normalizado es superior a 1,5. En contraste, los trabajos con cuatro o menos autores han recibido 1,6 citas como promedio y su impacto normalizado es de 1,0. Las medianas de este grupo son de 1 y 0,25 respectivamente. La prueba de Kruskal-Wallis indica que las diferencias detectadas son significativas ($X^2= 275,89$, $p < 0.001$).

## DISCUSIÓN Y CONCLUSIONES

El análisis cuantitativo de los trabajos españoles de investigación en 20 áreas temáticas de las ciencias sociales, los datos evolutivos y la comparación con cifras internacionales descartan una inflación sistemática del número de autores. No obstante, se observan diferencias significativas entre categorías temáticas. Por ejemplo, en las categorías de Antropología y de Educación especial, se rebasa el umbral de 4 autores a partir de los cuales se penaliza las contribuciones de los autores españoles. La colaboración entre paleoantropólogos, biólogos evolutivos y especialistas en genética de poblaciones y paleogenética es común en la investigación antropológica actual. En el caso de la Educación especial, casi la cuarta parte de los trabajos publicados en el último año del periodo contienen contribuciones de grupos de investigación psicológica o psiquiátrica, lo que aproxima la investigación en esta categoría a la que se produce en ciencias del comportamiento, cuyas listas de firmantes también son numerosas. Por otra parte, aunque las restantes áreas presentan un tamaño de los equipos inferior, la comparación entre los años extremos del periodo sugiere que la tendencia hacia el crecimiento del número de autores es general.

El número de autores depende de la colaboración institucional. En el caso de la colaboración internacional, los autores extranjeros igualan o superan a los nacionales. Mal se puede penalizar a los grupos españoles por un fenómeno, el de la colaboración internacional, cuando ésta supone una ventaja epistémica y dota a los trabajos de mayor influencia e impacto (van Raan, 1998). En relación con el impacto, hemos observado que los trabajos cuyo número de autores supera el umbral son aquellos más citados y con mayor impacto normalizado. Las diferencias no sólo son significativas, sino muy amplias en algunas de las categorías temáticas. Este resultado es coherente con la observación común de que la colaboración favorece la realización de trabajos con mayor impacto y más influencia. Si los trabajos con mayor número de autores son aquellos con mayor nivel de internacionalización y mayor influencia cabe cuestionarse hasta qué punto es lógico penalizar a los equipos que sobrepasan un umbral que parece arbitrario.

En el contexto global, que la colaboración internacional es cada vez más la norma que la excepción. De hecho, el promedio de autores por artículo casi se ha duplicado entre 1955 y 2000, pasando de 1,9 a 3,5 (Wuchty y otros, 2007). El aumento es mayor en las ciencias sociales y se observa también en áreas donde tradicionalmente no ha sido necesaria la participación de grandes equipos de investigación (Henriksen, 2016). Asimismo, se ha evidenciado una relación positiva entre el número de autores y una mayor productividad de los grupos, así como una mayor tasa citación (Franceschet y Costantini, 2010) y una mayor obtención de fondos (Rosenzweig y otros, 2008; Defazio y otros, 2009).

Establecer políticas y criterios que pretenden evitar malas prácticas en lugar de promover buena ciencia, ponen en cuestión no la actitud del investigador español a la hora de colaborar y firmar trabajos con otros autores, sino la finalidad última de estos procesos evaluativos. La tendencia hacia una explicitación mayor de qué se acepta y qué no dentro de unos supuestos procesos de revisión por pares en aras de una mayor 'transparencia' (Derrick y Pavone, 2013), ignora principios básicos sobre el uso de los indicadores bibliométricos para la evaluación (Hicks y otros, 2015). Además, refuerza la función de control reflejo de una cultura administrativa que desconfía de sus trabajadores y cuya función es monitorizar y verificar que lo que se presenta es real en lugar de implantar una evaluación que persiga unos objetivos nacionales que contribuyan a mejorar el sistema científico nacional (Rafols y otros, 2016).

## Agradecimientos
Nicolás Robinson-Garcia disfruta actualmente de un contrato postdoctoral Juan de la Cierva-Formación financiado por el Ministerio de Economía y Competitividad. Los autores agradecen a Pablo D'Este e Ismael Ràfols (INGENIO) sus comentarios en conversaciones informales y a François Perruchas (INGENIO) por su ayuda en el procesamiento de los datos.

## Acknowledgements
Nicolás Robinson-Garcia is currently supported by a postdoctoral Juan de la Cierva-Formación grant from the Spanish Ministry of Economy and Competitiveness. The authors thank Pablo D'Este and Ismael Ràfols (INGENIO) for informal conversations on previous versions of this manuscript and François Perruchas (INGENIO) for his assistance on data processing.

## REFERENCIAS